\documentclass[10pt,journal,onecolumn]{IEEEtran}
\IEEEoverridecommandlockouts
\usepackage{cite}
\usepackage{amsmath,amssymb,amsfonts}
\usepackage[table,xcdraw]{xcolor}
\usepackage{graphicx}
\usepackage{lscape}
\usepackage{textcomp}
\usepackage{xcolor}
\usepackage[bookmarks=false]{hyperref}
\usepackage[noend]{algpseudocode}
\usepackage[linesnumbered,ruled,vlined]{algorithm2e}
\usepackage{amsmath,amsfonts}

\SetCommentSty{mycommfont}
\SetKwInput{KwInput}{Input}                
\SetKwInput{KwOutput}{Output}              
\def\BibTeX{{\rm B\kern-.05em{\sc i\kern-.025em b}\kern-.08em
    T\kern-.1667em\lower.7ex\hbox{E}\kern-.125emX}}

\algblock{Input}{EndInput}
\algnotext{EndInput}
\algblock{Output}{EndOutput}
\algnotext{EndOutput}

    \newcommand*{\affaddr}[1]{#1} 
\newcommand*{\affmark}[1][*]{\textsuperscript{#1}}
\newcommand*{\email}[1]{\textit{#1}}

\begin{document}

\title{Detection of Coincidentally Correct Test Cases through Random Forests \\
}

\author{%
Shuvalaxmi Dass\affmark[1], Xiaozhen Xue\affmark[2], Akbar Siami Namin\affmark[1]\\
\affaddr{\affmark[1]Department of Computer Science, Texas Tech University}
\affaddr{\affmark[2]Adobe Inc.,} \\
\email{shuva93.dass@ttu.edu; xxuettu@gmail.com; akbar.namin@ttu.edu}\\
}

\maketitle

\begin{abstract}
The performance of coverage-based fault localization  greatly depends on the quality of test cases being executed. These test cases execute some lines of the given program and determine whether the underlying tests are passed or failed. In particular, some test cases may be well-behaved (i.e., passed) while executing faulty statements. These test cases, also known as coincidentally correct test cases, may negatively influence the performance of the spectra-based fault localization and thus be less helpful as a tool for the purpose of automated debugging. In other words, the involvement of these coincidentally correct test cases may introduce noises to the fault localization computation and thus cause in divergence of effectively localizing the location of possible bugs in the given code. 
In this paper, we propose a hybrid approach of ensemble learning combined with a supervised learning algorithm namely, Random Forests (RF) for the purpose of correctly identifying test cases that are mislabeled to be the passing test cases. A cost-effective analysis of flipping the test status or trimming (i.e., eliminating from the computation) the coincidental correct test cases is also reported.

\end{abstract}


\section{Introduction}
\label{sec:intro}

Many studies and research have been conducted on improving the performance of coverage-based fault localization in spotting faults' location in programs. The precision of these automated debugging techniques often get deteriorated by the presence of ``{\it coincidentally  correct test cases,}'' which are test cases that do not expose the faulty behavior of the program, even though they are exercising the faulty portion of the given code. It poses as a hurdle to the testers, who are misled into effectively localizing the faulty code.

As an instance of the traditional classification problem, the identification of coincidentally correct test cases can be modeled using machine learning techniques \cite{6899222}. Hence, the problem formulation of coincidentally correct cases through machine learning techniques can be viewed as those test cases that are in fact failing but are mistakenly labeled into passing test cases. Hence, the name given ``coincidental'' to the test case in question appears to be passing. In short, the task at hand is to correctly identify the mislabeled passing test case and then either 1) re-label them into failing ones (i.e., flipping), or 2) completely ignore their contributions to the fault localization computation (i.e., trimming).

This paper proposes an ensemble-based random forests approach to identify coincidentally correct test cases. There are some other machine learning-based approaches to this problem (i.e., \cite{6899222}) where a classifier (e.g., Support Vector Machine (SVM)) is utilized to re-label test cases. However, what makes ensemble random forests a great classifier is its ability in building several decision trees and then taking the majority votes for making the final classification decision. Given the strong evidence observed in similar studies \cite{ICMLA2016} and due to ensemble-learning nature of random forests, it is expected that this machine learning technique outperforms other classification techniques \cite{RF/SVM}.

To measure the effectiveness of this approach, we conducted two types of experiments/strategies on the coincidentally correct test cases identified by random forests model: 1) trimming,  and 2) flipping followed by examining and comparing their effects on fault localization in terms of cost analysis. According to the results, the trimming strategy, by which the coincidentally correct test cases are removed 1) one at a time, and 2) all at once, enhances the performance of fault localization; whereas, the flipping strategy, did not exhibit any improvement over some other classification algorithms. Moreover, the cost analysis results show that the flipping and trimming coincidentally correct test cases all at once performed better than doing them one at a time. The results also show that the proposed ensemble-based random forests technique performed better as compared to the technique that used SVM in terms of cost analysis. This paper makes the following key contributions:

 \begin{enumerate}
     \item Introduce an ensemble-based random forests classifier to identify test cases with inaccurate or noisy test results. 
     \item Conduct a performance analysis on the improvement achieved using ensemble-based random forests over other classification techniques. 
     \item Report the results of a comparison of two strategies in dealing with coincidentally correct test cases (i.e., trimming/flipping) in terms of cost metrics.
 \end{enumerate}

Section \ref{sec:relatedwork} reviews the related works. The background of machine learning algorithms used in this paper is presented in Section \ref{sec:background}. Section \ref{sec:methodology} describes the methodology of employing random forests for mislabeled test cases. Section \ref{sec:example} illustrates an example to demonstrate the methodology. Section \ref{sec:experiment} presents the results and evaluations. Section \ref{sec:conclusion} concludes the paper and provides some hints on future work.

\section{Related Work}
\label{sec:relatedwork}

There are several research work and studies regarding identification of coincidentally correct test cases. Li et al.\ \cite{7780197} proposed a machine learning-based fuzzy classification technique to identify coincidentally correct test cases followed by the application of the KNN algorithm to identify the remaining passed test cases. Feyzi and Parsa \cite{F} proposed a SVM-based algorithm with a customized kernel function to  improve fault localization effectiveness by removing the impact of coincidentally correct test cases. Xue et al.\ \cite{6899222} proposed a technique combining support vector machine and ensemble learning to detect mislabeled test cases followed by flipping and trimming their test status in order to improve the performance of fault localization. Pang et al. \cite{PangXN13} also proposed identifying effective test cases through K-Means Clustering for enhancing regression testing. 
 
Zhang et al.\ \cite{Z} introduced a test classification technique to assign labels to the unlabeled test cases based on the information gathered from their execution traces. This enables making those newly labeled test cases be utilized in localizing faults thereby improving on its effectiveness. 

Patel \cite{Pat} empirically evaluated four techniques to mitigate the coincidental correctness in testing and debugging. The first technique is called Interlocutory Testing, which is a framework in which test oracles are developed. The second technique called the Interlocutory Metamorphic Testing that mitigates the impact of coincidental correctness when combined with Interlocutory Testing. The other two techniques, Interlocutory Testing and the Interlocutory Spectrum-based Fault Localization, both work towards alleviating the effects of coincidental correctness on fault localization.

\section{BACKGROUND}
\label{sec:background}


\subsection{Ensemble Learning}
Ensemble learning is a machine learning technique  where multiple learners are trained to address the same problem. It is different from other ordinary machine learning approaches in terms of the different number of models it uses to train and then combining them in order to make the final decision. In this technique, different models are employed to train on the multiple disjoint data subsets obtained from the original dataset. Each model, after being trained, makes its own prediction when applied on the test data. Each prediction (i.e., vote) is then aggregated along with other predictions made by other models into one final prediction. This process and the final decision made is called a {\it majority voting}. This model is utilized in random forests to identify the coincidentally correct test cases, which are mislabeled into passing test cases.
 
\subsection{Random Forests Classification}
\begin{figure}
\centering
\includegraphics[width=\linewidth]{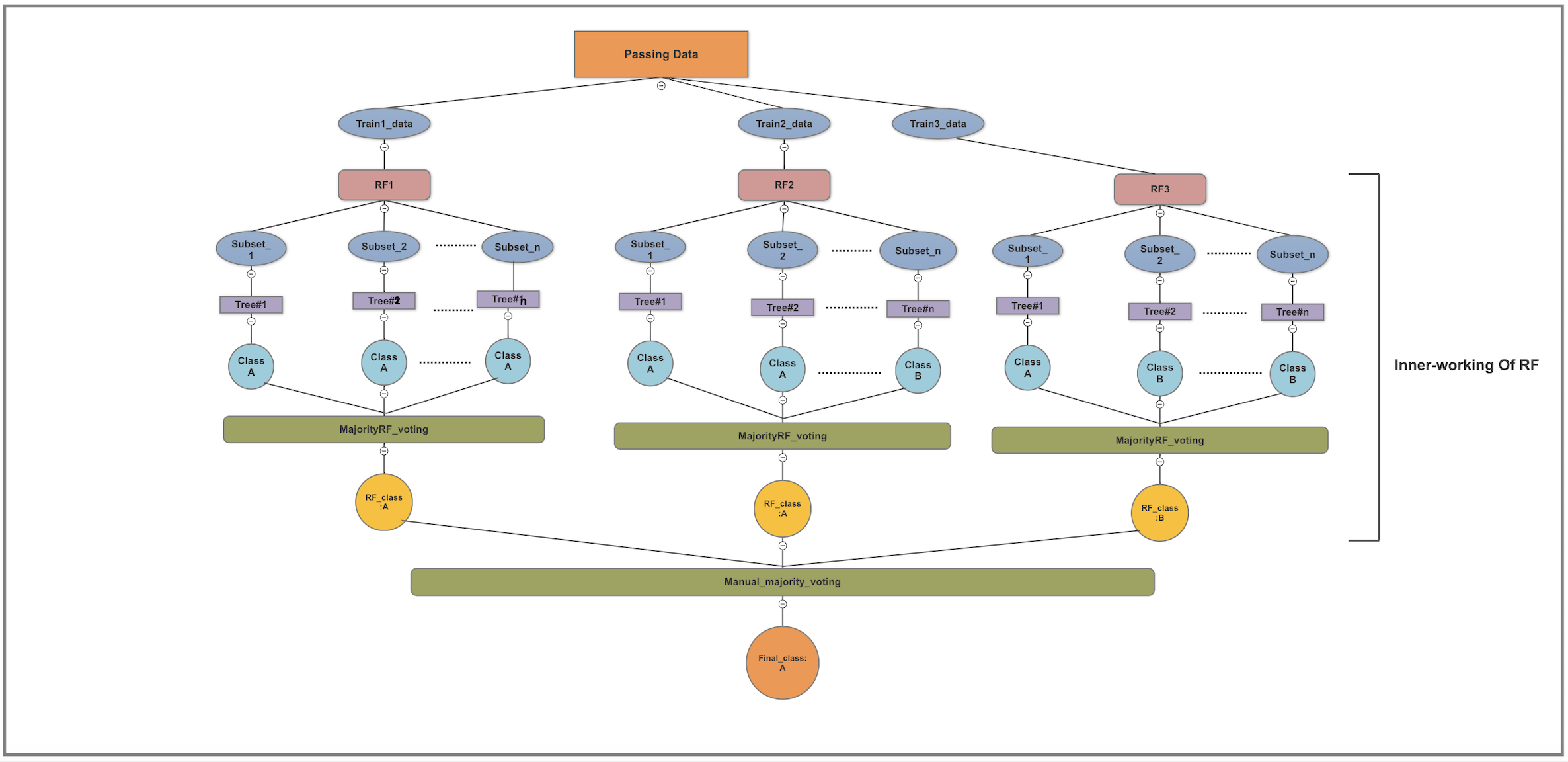}
\caption{\label{fig:RF} A representation of random forest embedded in ensemble learning.}
\vspace{-0.2in}
\end{figure}

In this paper, we made use of the random forests classification instead of employing some other types of classifiers with the goal of achieving better results for detecting coincidentally correct test cases. The random forests classifiers are supervised ensemble-learning models used for classification and regression. The idea behind ensemble learning models is to make use of  multiple machine learning models and aggregate them to obtain overall better performance.
The rationale behind this is every one of the models utilized is frail when used individually, yet solid when put together in an ensemble. 
A typical random forests classifier consists of multiple ``{\it weak}'' decision trees whose outputs when combined result in a ``{\it strong}'' ensemble. Figure \ref{fig:RF} show a graphical representation of how random forests works. 

\subsection{Principle Component Analysis}
Principal component analysis (PCA) is a dimensionality reduction technique of a dataset.
Its basic functionality is to create principal components, which are basically independent features that are a combination of a number of dependent features. The primary principal component being the one that possess the maximum variation as possible in the dataset and the components followed represent the significant portion of the remaining inconstancy and variance as possible. 

In our problem domain, we deal with large program files that contain a huge number of code lines. These individual statements act as features (i.e., dimensions) which can grow in size depending on the size of the program and consequently make the computation complex. Hence after much experimentation, we decided to reduce the dimensionality of our dataset to 60 percent of the actual number of dimensions, which would represent the best features. We applied Principal Component Analysis (PCA) to our dataset setting attribute {\tt n\_components} $= 0.6$ that controls the number of best attributes needed to build a better classifier.

\section{Methodology}
\label{sec:methodology}

This section presents the methodology of identifying mislabeled test cases through random forests.

\subsection{Notations}
The overall motive of the technique is to predict the coincidentally correct test cases out of the pool of passing test cases devised for a faulty program. To begin with, we introduce some notations for  representation and formulation purposes.

\begin{itemize}
\item $PT = \{ pt_{1}, pt_{2},....pt_{i} \}$: Passing test cases where $pt_{i}$ represents the $i$-th passing test case.

\item $FT = \{ ft_{1}, ft_{2},....ft_{j} \}$: Failing test cases where $ft_{j}$ represents the $j$-th failing test case.

\item $CT = \{ ct_{1}, ct_{2},....ct_{l} \}$: Coincidentally correct test cases. $ct_{l}$ is the $l$-th coincidentally correct test case. 
\end{itemize}

\subsection{Partitioning Dataset into Train/Test Sets}

\begin{figure}
\centering
\includegraphics[width=\linewidth]{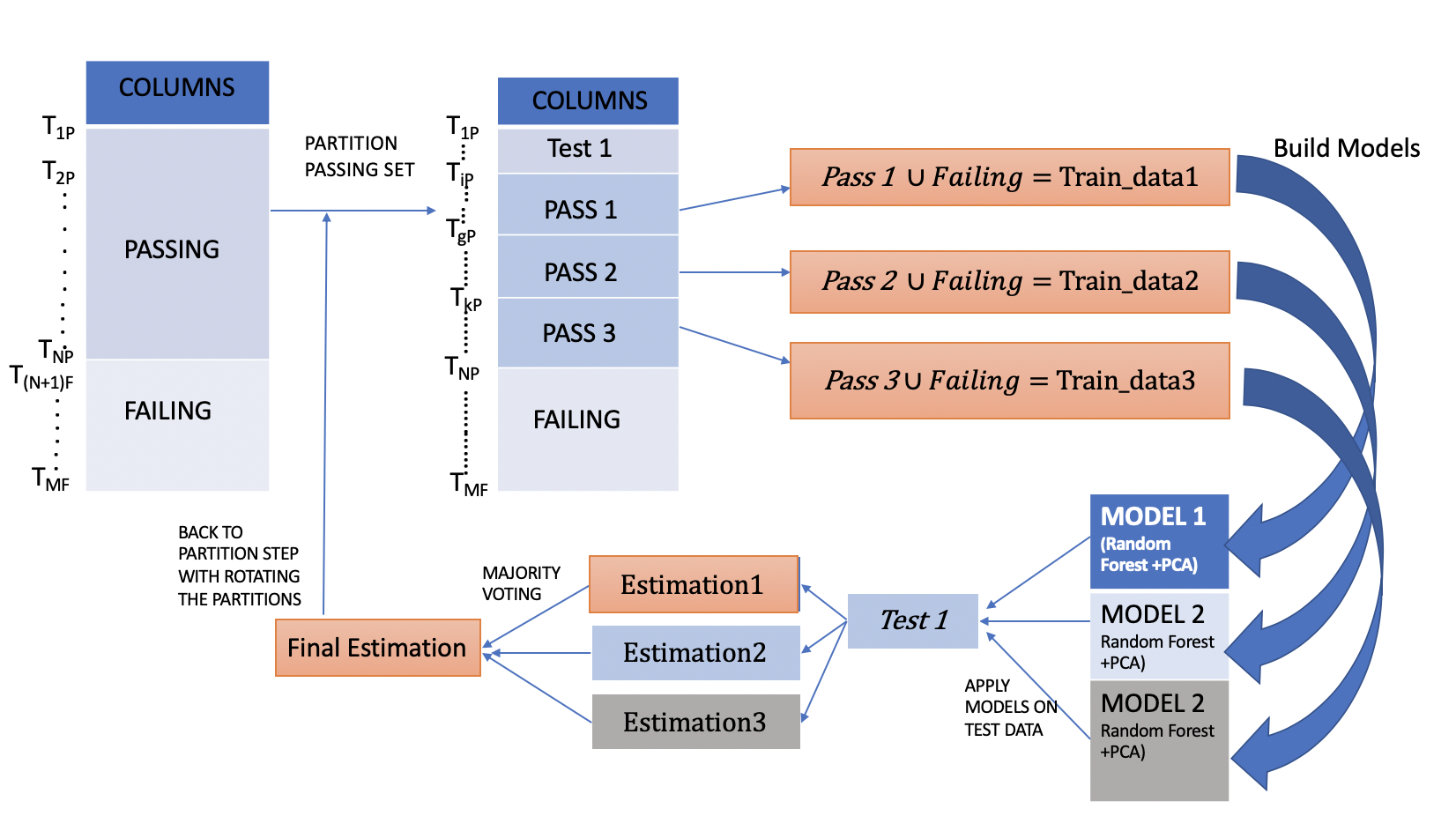}
\caption{\label{fig:1}Partitioning procedure.}
\label{fig:Partitioning}
\vspace{-0.2in}
\end{figure}

The unit test cases corresponding to each faulty program are first divided into two sets: failing and passing. The passing data set itself is partitioned into $m$ subsets, out of which one subset corresponds to the test dataset ($Test_{1}$) and the remaining sets $m-1=l$ subsets (i.e., $Train_{1}$, $Train_{2}$, ..., $Train_{k}$)  correspond to the training set. The failing test cases are then added to each training set. In other  words, each of the training set is a combination of {\it selected} passing test cases and {\it all} failing test cases. Each of these training sets are then used to train our random forests-based model. The procedure of partitioning passing data to build each subset $Train_{i}$, for $i=1$ to $k$ is depicted in Figure \ref{fig:Partitioning}.

The train dataset will comprise of both failing and passing test cases. Hence, each $Train_{i}$ will include subset of data from passing test case ($PT$) dataset and all of the data from the failing test case ($FT$) dataset. It is important to note that the number of passing test cases need to be (and usually it is) more than the failing ones. We included the entire failing test cases in each of k subsets, as it was generally less than the passing ones. 
As a result, each subset $Train_{i}$ comprised of randomly selected a small subset of passing test cases:

\[ Pt_{i} \subset PT  \] 

and all of the failing test cases (FT). The Train set is:

 \[  Train_{i} = \big \{ Pt_{i} \subset PT \} \cup \big \{FT\}  \] 

In our faulty programs under test, we partitioned each passing test cases into four subsets out of which one set is the test set consisting of PT cases. The size of the test data is based on intuition and is generally taken less than that taken for the train sets. Whatever left in the passing dataset, is equally divided into three subsets where each part along with added $FT$ corresponds to one train sets. A random forests (RF) classifier is then trained based on these subsets. In total, three such sets were formed, and three RF classifiers were trained to do classification on the test set. The algorithm continues this entire procedure until all the passing test cases are processed one test set at a time.

\subsection{The Algorithm}

Algorithm \ref{alg:CC} describes the procedure for employing the random forests (RF) classifier to identify coincidentally correct test cases (CC) for a given faulty program.

\begin{algorithm}[]
\DontPrintSemicolon
  \KwInput{
           1) Passing Test Cases $PT = \{pt_{1},pt_{2},..pt_{N}\}$
           \newline
           2) Failing Test Cases $FT = \{ft_{1},ft_{2},..ft_{M}\}$
           \newline
           3) Random Forests Learning Algorithm: RF
           }
  
  \KwOutput{$CT$}  \tcp*{Coincidentally correct test cases}
  {Choose a random size  $K \leq N $
  \newline
  test = \{\} ; CT = \{\};}
  \newline
  
  \For {($i =0$, $i = N$, $i+=K$)}
    {
         \For {($j = i+1$ to $i+K $)}{
              $test.append(pt_{j})$
           }
        {Randomly divide the (N-K) passing test cases into 'p' equal partitions}
        \newline
         \For{$m = 1$ to p}
        {
    	    $ train_{m} = partitions_{m} \cup FT $
    	    \newline
    	    $ RF_{m}= L(train_{m}) $
    	    \newline
    	    $ Z_{m}= \{\} $ \tcp*{Initialization of each label pool}
        
             \For {each test data  $pt_{k}$  in test}
        { 
        \tcp*{label test data as P or F by RF}
        	$label_{k} = RF_{m}(pt_{k})$ 
        	\newline
        	$Z_{m} = Z_{m} \cup \{label_{k}\}$
        }
    }
  \For {each test data $pt$ in $test$}
    { 
            $ cc_{num} = 0 $ \tcp*{\# of CT test cases}
            $ ncc_{num} = 0 $ \tcp*{\# of non-CT test cases} 
            \For {$m = 1$ to p}
            { 
              \If {$Z_{m}$ labels $pt == coincidentally\_correct$}
              {
                $cc_{num}++$
              }
             \Else 
             {
               $ncc_{num}++$
             }
           }
       \If { $cc_{num} \geq  ncc_{num}$ }
          {
          $CT = CT \cup \{pt\}$  \tcp*{Majority Voting}
          }
      }
    } 
\caption{Identifying coincidentally correct test cases.}
\label{alg:CC}
\vspace{-0.05in}
\end{algorithm}

The input to the algorithm is a set of passing (i.e., $PT$) and failing (i.e., $FT$) test cases and the classifier embedded is random forests. The output is a set of coincidentally correct test cases (i.e., $CT$). The algorithm encompasses five steps:

\begin{enumerate}
 \item \textbf{Initialization (lines 2-4).} It initializes the test dataset from selecting   randomly chosen size $K$ unit test cases starting from beginning  from  passing test cases dataset.
 
 \item  \textbf{Portioning (lines 5-6).} It builds up three training sets (i.e., partitions) by merging the entire failing test cases with each of the partition. A random forests (RF) classifier is also trained on each of the training set. The L function represents the random forests learning module.A label pool set Z is also initialized for every partition to store the labelled test cases. 
 
 \item \textbf{Classification (lines 7-8).} The trained RF classifier is then used to label the status of each test case present in \textit{test} set  and then store them in the label test case set $Z$ corresponding to the partition/training data on which the RF classifier is trained on. 
 
\item \textbf{Ensemble (lines 9-16).} It counts the number of coincidentally and non-coincidentally correct labels for each test case that are produced by the implicit decision tress generated by the RF classifier. 

\item \textbf{Decision and Voting (lines 17-18).} If the number of coincidentally correct labels are more than non-coincidentally ones, then by majority voting, the corresponding test case is stored in the $CT$.
 
\end{enumerate}

\subsection{Metrics}
We measured the effectiveness of the machine learning algorithm, RL based on the \textit{fault localization cost} metric, which is calculated as the percentage of statements that must be examined before reaching the first faulty statement in a faulty program \cite{6899222}. Once we get the the coincidentally correct test case labels returned by the algorithm, we perform the two strategies: 1) Flipping and 2) Trimming on those cases, and then the fault localization procedure is re-performed to compare the cost of fault localization for each strategy to that of no strategy (original cost). Reduction in the cost from the original is indicative of the the algorithm performing well.  

\section{An Illustrative Example}
\label{sec:example}

\begin{figure}
\centering
\includegraphics[width=\linewidth]{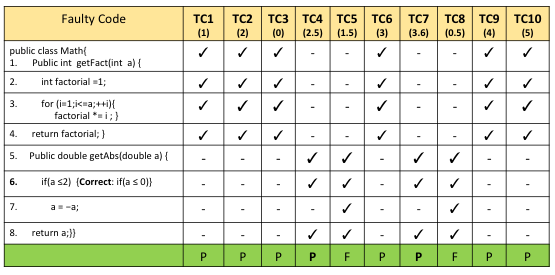}
\caption{\label{Table:example} An illustrative example.}
\vspace{-0.2in}
\end{figure}

This section presents an illustrative example to demonstrate the mechanic of the presented ensemble-based random forests. Consider a  faulty Java program of eight lines of code given in Figure \ref{Table:example}. This Java snippet consists of a class Math in which two methods are defined. The first method {\tt getFact()} takes in an integer number and computes the factorial value of a number. This class also implements another method namely, {\tt getAbs()}, which takes double value as a parameter. A fault is injected at line 6 where the logic condition in the ``$if$'' loop is incorrectly set to 'a $<=$2' instead of 'a$<=$0'. 

\begin{figure}
    \centering
    \includegraphics[width=\linewidth]{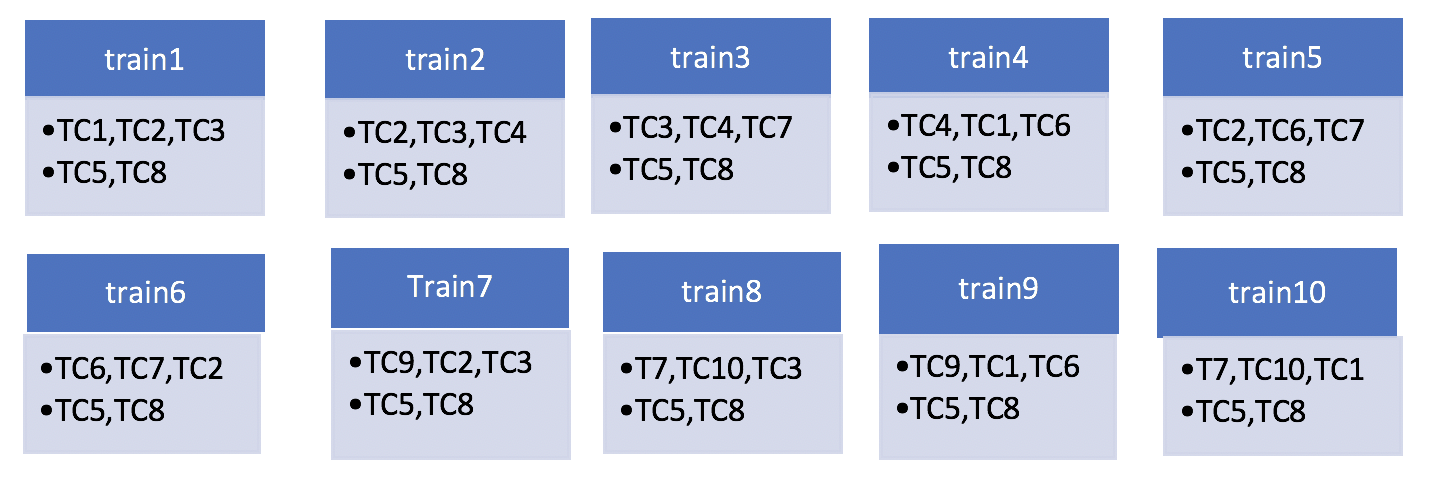}
    \caption{Training data sets.}
    \label{fig:ExampleDataSets}
    \vspace{-0.2in}
\end{figure}

The test pool consists of eight passing and two failing test cases. The passing test cases $TC4$ and $TC7$ both execute the faulty statement without exhibiting any faulty behavior. Hence, $TC4$ and $TC7$ are apparent \textit{coincidentally correct test cases}. The purpose is to spot such coincidentally correct test cases using ensemble-based random forests as a classifier. To begin with, the algorithm splits the passing test cases into N groups and then add all the failing test cases to each group. Here, we divided the test cases into 10 (i.e., $N = 10$) random groups . Each group is composed of three passing test cases (chosen in random) along with two failing test cases. Figure \ref{fig:ExampleDataSets} represents what each data set comprises of.

\begin{figure}[!t]
    \centering
    \includegraphics[width=\linewidth]{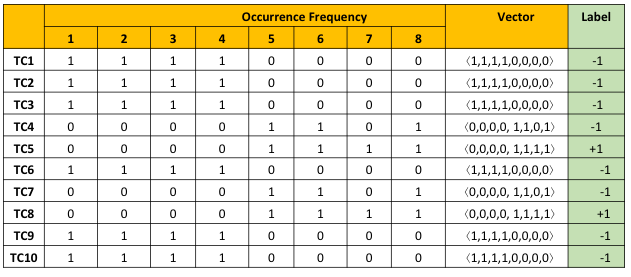}
    \caption{Vectors For test cases: '-1' for passing, '+1' for failing.}
    \label{fig:Prediction}
\vspace{-0.2in}
\end{figure}

\begin{figure}
    \centering
    \includegraphics[width=\linewidth]{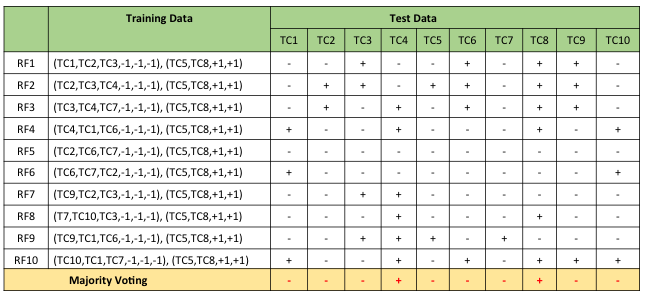}
    \caption{Majority Voting: '-1' for passing, '+1' for failing.}
    \label{fig:Examplevoting}
    \vspace{-0.12in}
\end{figure}

Using each training set, a single RF classifier is built and trained on each training set. In this example, ten such classifiers were built for predicting the test status of the passing test cases. Figure \ref{fig:Prediction} shows the prediction results of the 10 classifiers.  Then, the majority votes were taken for each passing test case as shown in in Figure \ref{fig:Examplevoting}. 

For example, in case of test $TC1$, 7 out of 10 classifiers labeled this test case as '-1'. Hence, the majority decision for $TC1$ was '$-1$' and therefore $TC1$ was labeled as a true passing test case and it was not a coincidentally correct test case. On the other hand, in case of $TC4$, 6 out of 10 predictions were '$+1$'. Therefore $TC4$ was labeled as a coincidentally correct test case.

\section{Experimental Evaluation}
\label{sec:experiment}

This section reports the performance of the introduced approach on improving fault localization. The improvement is measured through the reduction of debugging cost achieved through 1) no removal, 2) flipping, and 3) trimming of the coincidentally correct test cases identified by the methodology. 

\subsection{Subject Programs}

Table \ref{tab:my-table} lists the subject programs studied in this research. We obtained these widely-studied programs through the SIR repository (Software-artifact Infrastructure Repository).

\begin{table}[!t]
\caption{Subject programs and the number of coincidental correct test cases.}
\centering
\begin{tabular}{|c|c|c|c|c|}
\hline
\textbf{Programs} & \textbf{Version(Fault No.)} & \textbf{Combo1} & \textbf{Combo2} & \textbf{Combo3} \\ \hline
Jmeter & V1(F1)       & 3               & 3               & 3               \\ \hline
Jmeter & V3(F2)       & 2               & 4               & 1               \\ \hline
Nano & V1(F1)         & 31              & 4               & 20                 \\ \hline
Nano & V1(F2)         & 32              & 24              & 25              \\ \hline
Nano & V1(F4)         & 17              & 16              & 15                 \\ \hline
Xml & V2(F3)          & 7               & 8               & 7               \\ \hline
Jtop & V3           & 1               & 1               & 1               \\ \hline
\end{tabular}
\vspace{-0.3in}
\label{tab:my-table}
\end{table}

\subsection{Data Collection}
For each faulty program used for testing purposes, there are three files namely: Combo1, Combo2, and Combo3 each pertaining to  1, 2 and 3 number of statements (i.e., execution trace)  that the test cases execute at a time, respectively. These statements act as features. In other words, Combo1, Combo2, and Combo3 comprise a combination of single, double and triple statements as features that are executed by test cases at once, respectively.

We built three separate data sets each corresponding to the combo files and performed the partitioning as  discussed in the previous section on each dataset to eventually identify coincidentally correct test cases present in each combo files.

{\it Building Feature Vectors.} Coming to the contents of dataset part, where features represent the statements executed by each test case, the value under each feature represents the count of number of times a test case is executing that statement. For instance, consider a faulty program with seven lines of code each executable in nature and suppose an execution trace of a test case $t_1$ in the form of a vector that looks like 

\[ ExecutionTrace_{t_{1}}: \big \langle 1, 2, 5, 6, 5, 2 \big \rangle \]

where each value represents statement number executed by that test case. Each statement number of that test case would have a vector of  values of the form that can be represented as:

\[ FeatureVector_{t_{1}}: \big \langle 0, 2, 2, 1, 2, 0 \big \rangle \] 

where each value represents the count of times the test case executed the underlying statement number.

\subsection{Partitioning Passing and Failing Test Cases}

We partitioned total number of \textit{N} passing test cases into 1) one test data file containing K = 10 (randomly chosen) number of passing test cases for prediction , 2) three train dataset files each containing $(N-10)/3$ number of test cases, 3) add all the failing test cases from the failing dataset file into each of the three train files in order to build and train three models. The models then are used to obtain three different classifications for a single passing test case. Out of these three predicted labels, the majority voting would be the final status (i.e., passing or failing) of the underlying test case.
We recursively followed the above steps until we cover all the test cases in the passing dataset file and in the end, we obtain the final majority voting.

\begin{table*}[!t]
\caption{Cost Analysis with flipping.}
\centering
\begin{tabular}{|l|c|c|c|c|c|c|c|}
\hline
\textbf{Flipping Cost} & \multicolumn{3}{c|}{\textbf{One a Time (Variant 1)}} & \multicolumn{3}{c|}{\textbf{All at Once (Variant 2)}} & \textbf{Original Cost}                   \\ \hline
\textbf{Programs}      & \textbf{combo1}         & \textbf{combo2}         & \textbf{combo3}         & \textbf{combo1}          & \textbf{combo2}         & \textbf{combo3}         &                                          \\ \hline
Jmeter V3F2            & 0.022                   & \{0.022,0.933\}         & 0.022                   & 0.022                    & 0.022                   & 0.022                   & { \textbf{0.022}}    \\ \hline
Jmeter V1F1            & \{0.08, 0.06\}          & \{0.08, 0.06\}           & \{0.08, 0.06\}           & {\textbf{0.06}}               & 0.06                    & 0.06                    & { \textbf{0.16}}     \\ \hline
Nano V1F1              & \{0.093, 0.140\}      & \{0.093, 0.140\}      & 0.047                 & 0.062                   & 0.078                & 0.047                 & { \textbf{0.171}} \\ \hline
Nano V1F2              & 0.562                  & 0.562                  & 0.562                  & 0.218                  & 0.262                  & 0.265                  & { \textbf{0.562}}   \\ \hline
Nano V1F4              & 0.70312                 & 0.703                 & {\textbf{0.509}}            & 0.703                  & 0.703                 & 0.509                  & { \textbf{0.703}}  \\ \hline
Xml V2F3               & \{0.058, 0.098\}     & \{0.058, 0.098\}     & \{0.058, 0.098\}     & 0.098                  & 0.058                 & 0.098                 & 0.058   \\ 
\hline
Jtop V3                & 0.172                & 0.172                & 0.172                & 0.172                 & 0.172                & 0.172                & { \textbf{0.172}} \\ \hline
\end{tabular}
\label{tab:my-table-flipping}
\end{table*}

\begin{table*}[!t]
\caption{Cost analysis with trimming.}
\centering
\begin{tabular}{|l|c|c|c|c|c|c|c|}
\hline
\textbf{Trimming Cost} & \multicolumn{3}{c|}{\textbf{One a Time (Variant 1)}} & \multicolumn{3}{c|}{\textbf{All at Once (Variant 2)}} & \textbf{Original Cost}                   \\ \hline
\textbf{Programs}      & \textbf{combo1}           & \textbf{combo2}        & \textbf{combo3}        & \textbf{combo1}          & \textbf{combo2}         & \textbf{combo3}         &                                          \\ \hline
Jmeter V3F2            & 0.022                     & \{0.022,0.9\}          & 0.022                  & 0.81                     & 0.022                   & 0.022                   & { \textbf{0.022}}    \\ \hline
Jmeter V1F1            & 0.16                      & 0.16                   & 0.16                   & {\textbf{0.08}}               & 0.08                    & 0.08                    & { \textbf{0.16}}     \\ \hline
Nano V1F1              & 0.171                    & 0.171                 & 0.156                & {\textbf{0.156}}            & 0.171                  & 0.156                 & { \textbf{0.171}} \\ \hline
Nano V1F2              & 0.562                    & 0.562                 & 0.562                 & {\textbf{0.218}}            & 0.218                 & 0.218                 & { \textbf{0.562}}   \\ \hline
Nano V1F4              & 0.70312                   & 0.703                & {\textbf{0.478}}           & 0.703                  & 0.70312                 & 0.478                  & { \textbf{0.703}}  \\ \hline
Xml V2F3               & \{0.058, 0.176\}         & 0.058                & 0.058                & 0.176                    & 0.058                 & 0.058                 & 0.058                                  \\ \hline
Jtop V3                & 0.172                  & 0.172               & 0.172               & 0.172                 & 0.172                & 0.172                & { \textbf{0.172}} \\ \hline
\end{tabular}
\label{tab:my-table-trimming}
\end{table*}

Table \ref{tab:my-table} also lists the number of coincidentally correct test cases identified in each of the three combo files corresponding to the faulty programs under test. We identified the coincidental correct test cases by majority voting which were then subjected to flipping and trimming procedure. 

\subsection{Fault Localization Cost Analysis}

We applied the aforementioned techniques (i.e., flipping and trimming) to our dataset and identified the final list of coincidentally correct test cases for each program. We developed and ran our python script to find the cost of localizing a fault first by trimming and then by flipping the coincidentally correct test cases and report the results.

We ran python scripts which took as input the trace, test status and the instrumentation files for each program under test to calculate the cost of localizing fault. Two variants of cost of fault localization were evaluated to observe which one is offering the best results. The results also were compared with the original cost, which is the cost calculated when no changes were done to the test suite. In the first variant, we calculated the cost of flipping/trimming the coincidentally correct test cases one at a time; whereas, in the second variant we calculated the cost after trimming/flipping all the identified coincidentally correct test cases at once.

Tables \ref{tab:my-table-flipping} and \ref{tab:my-table-trimming} report the cost analysis of flipping and trimming procedure. We observe that for most of the subject programs, the second variant, i.e, doing the changes to the test suite all at once exhibited lesser than or equal cost value to the original cost as compared to the first variant.

\subsection{Flipping and Trimming Cost Analysis}
In Table \ref{tab:my-table-flipping}, some of the combo files under the first cost variant have two cost values representing the values in which the coincidentally correct test cases fluctuated in. The values marked in bold indicated the corresponding subject programs incurred lesser or equal cost of fault localization when its coincidentally correct test cases were flipped all at once. A similar trend was seen in the case of trimming in Table \ref{tab:my-table-trimming}. 

We observe that adopting our technique of ensemble-based random forests mechanism can prove to be beneficial in terms of fault localization cost as it is pretty evident from the results reported that when the cost-evaluating python scripts ran on the test suite of the subject programs under test in which the RF-identified coincidentally correct test cases' status was flipped/trimmed, it gave better(read lesser) or equal cost results thereby making this mechanism a useful tool in reducing the expense incurred for fault localization.

\section{conclusion and Future Work}
\label{sec:conclusion}

This paper proposed an random forests-based ensemble technique to identify coincidentally correct test cases. The motive behind this proposal was to mitigate the high cost incurred in the localizing fault in a faulty program due to the  presence of coincidentally correct test cases. As a result, two strategies, namely, flipping and trimming were discussed and their performance in terms of incurred fault localization were analyzed. The results show that the proposed technique fared well in achieving the motive. There exist some other avenues in the field of machine learning that can be explored to work towards attaining this goal. 
In this paper, we worked on subject programs available in the SIR repository. It is important to replicate the experiment reported here on a larger set of programs (e.g. Defects4J \cite{defects4J}) so to observe the true benefit of ensemble learning approaches to this problem.  Moreover, in order to choose the best value for \textit{K}, the model might need some exploration on the optimization aspects of the problem and particularly performing some uncertainty  reasoning \cite{NaminS10}.

\section*{Acknowledgment}
This research work is supported in part by National Science Foundation under Grant No: 1821560.

\bibliography{References}
\bibliographystyle{plain}
\end{document}